\documentclass[journal,10pt]{IEEEtran}
\usepackage{cite}
\usepackage{amsmath,amssymb,amsfonts}
\usepackage{algorithmic}
\usepackage{graphicx}
\usepackage{amsthm}
\usepackage{epsfig}
\usepackage{color}
\usepackage{subcaption}
\usepackage{lettrine}
\usepackage{float}
\usepackage{lipsum}
\usepackage{bm}
\usepackage{mathtools}
\usepackage{bbm}
\usepackage{suffix}
\usepackage[T1]{fontenc}
\usepackage[colorlinks]{hyperref} 
\usepackage[nameinlink,noabbrev]{cleveref}
\usepackage{textcomp}
\usepackage{xcolor}
\usepackage[def-file=diffcoeff-doc]{diffcoeff}
\usepackage{stackengine}
\usepackage[ruled,norelsize]{algorithm2e}

\usepackage{textcomp}
\def\BibTeX{{\rm B\kern-.05em{\sc i\kern-.025em b}\kern-.08em
    T\kern-.1667em\lower.7ex\hbox{E}\kern-.125emX}}

\newtheorem{lemma}{Lemma}

\newcommand{\E}{\ensuremath{\mathbb E}}

\def \treq {\stackrel{\tiny \Delta}{=}}

\makeatletter
\def\@seccntformat#1{\@ifundefined{#1@cntformat}%
	{\csname the#1\endcsname\quad}
	{\csname #1@cntformat\endcsname}
	}
\newcommand{\removelatexerror}{\let\@latex@error\@gobble}
\makeatother

\begin{document}
\bstctlcite{IEEEexample:BSTcontrol}

\title{Joint Trajectory and Power Allocation Design for Secure Artificial Noise aided UAV Communications}

\author{
\IEEEauthorblockN{
Milad Tatar Mamaghani\IEEEauthorrefmark{1}, Yi Hong\IEEEauthorrefmark{1}, ~\IEEEmembership{Senior Member,~IEEE}}
		
\IEEEauthorblockA{
\IEEEauthorrefmark{1}Electrical and Computer Systems Engineering Department, Monash University, Melbourne, Australia}
}

\date{}
\markboth{}{}
\maketitle

\begin{abstract}

This paper investigates an average secrecy rate (ASR) maximization problem for an unmanned aerial vehicle (UAV) enabled wireless communication system, wherein a UAV is employed to deliver confidential information to a ground destination in the presence of a terrestrial passive eavesdropper. By employing an artificial noise (AN) injection based secure two-phase transmission protocol, we aim at jointly optimizing  the UAV’s trajectory, network transmission power, and AN power allocation over a given time horizon to enhance the ASR performance. Specifically, we divide the original non-convex problem into four subproblems, and propose a successive convex approximation based efficient iterative algorithm to solve it suboptimally with guaranteed convergence. Simulation results demonstrate significant security advantages of our designed scheme over other known benchmarks, particularly for stringent flight durations.

\end{abstract}

\begin{IEEEkeywords}
UAV communications, physical layer security, artificial noise injection, convex optimization.
\end{IEEEkeywords}

\section{Introduction}
\lettrine[lines=2]{R}{ecently}, unmanned aerial vehicle (UAV) aided wireless communications have attracted a great deal of research interests. 
UAV has unique attributes such as flexible deployment, dominant  line-of-sight (LoS) air-ground (AG) channel, and controlled mobility in three-dimensional (3D) space. As a result, UAV can act as either a flying base station (BS), aerial radio access point, or aerial relay via AG links to expand coverage, provide seamless connectivity, and support high-rate communications. However, the open nature of AG links inevitably makes such systems vulnerable to eavesdropping attacks, and hence, safeguarding UAV communication is of significant importance \cite{sun2019physical}. 

Physical layer security (PLS) techniques have been intensively studied in the past decade for terrestrial wireless communications with fixed or quasi-stationary nodes against eavesdropping (see \cite{Hamamreh2018PLS, Mamaghani2018sec}, and references therein). In the recent few years, a large number of research studies have been viewed UAV communications from the PLS perspective (see \cite{Wu2019Saf} and  references therein). Particularly, Zhang \emph{et al.} in \cite{zhang2019securing} have considered a three-node scenario and investigated UAV's trajectory design and transmission power control for both downlink and uplink UAV communications in the presence of a passive eavesdropper. Robust trajectory and power design of UAV communications against multiple ground eavesdroppers with imperfect locations has been explored in \cite{cui2018robust}. 
In \cite{Wang2017Mob}, the authors have investigated a secrecy rate maximization problem of the UAV-mobile relaying in the presence of an adversary. This work has been later extended by considering the UAV's trajectory design in \cite{Wang2018Joi} and the energy efficiency design in \cite{Xiao2018Sec}. 
The authors in \cite{Milad2019UAVAccess} proposed an energy-efficient low-altitude secure UAV relaying scheme with destination-assisted jamming and energy harvesting. Leveraging the technique of employing an additional UAV  as a mobile jammer has been studied in \cite{li2018uav} to improve the secrecy of  ground wiretap channel, and in \cite{cai2020joint, Mamaghani2020UAVJmr, Li2019Coo}, to enhance the security for the UAV-BS scenario.

On the other hand, artificial noise (AN) based PLS, originally proposed in \cite{GoelAN}, is a popular technique that can efficiently combat eavesdropping for terrestrial communication systems (e.g., \cite{LiuAN} for multiple-antenna systems, and \cite{He2017AN} for a single-antenna system). In \cite{He2017AN}, a two-phase transmission scheme was proposed in the presence of a passive eavesdropper: in phase I, the legitimate receiver broadcasts AN, and in phase II, the transmitter forwards the received signal from Phase I along with the information signal to the legitimate receiver. Then the receiver removes the AN and recovers the information, while the wiretap link can be degraded due to AN injection. To the best of our knowledge, such a technique has never been examined for UAV communications considering associated new degrees of freedom, e.g., AG links and mobility.

In this paper, motivated by the abovementioned works, we propose a secure UAV communication scheme by adapting a single-antenna AN-injection technique to UAV-enabled wireless communications. Different from \cite{He2017AN} that assumes static and fixed network nodes, our paper exploits the UAV's high-maneuverability and investigates secure UAV communications to maximize achievable average secrecy rate (ASR). 
Our work is also different from \cite{zhang2019securing}, since we employ a secure communication protocol via AN-assisted resource allocation rather than simple direct transmission. Overall, the central innovation of this work lies in the joint design of UAV's trajectory and system power allocation for the proposed scheme, which is also capable of achieving \textit{perfect secrecy} \cite{Hamamreh2018PLS}. Our contributions are summarized below.  
\begin{itemize}
\item 
Since the considered ASR maximization problem is intractable, we split this problem into subproblems according to the block coordinate descent (BCD) algorithm, solving each separately, as a compromise approach.
\item
 We then propose a fast converging and computationally efficient iterative algorithm based on the successive convex approximation (SCA) method, to find a local optima with guaranteed convergence.
 \item
 For the proposed algorithm, we conduct convergence and complexity analysis, and show that the solution can be obtained in polynomial time order, making it applicable to UAV-based scenarios.
 \item
  By simulations, we demonstrate that our proposed algorithm can significantly improve ASR and achieve a notably higher secrecy performance compared to other benchmarks.
\end{itemize}

\section{System model and problem formulation} \label{SysMod}
\begin{figure}[t]
\centerline{\includegraphics[width=0.5\columnwidth]{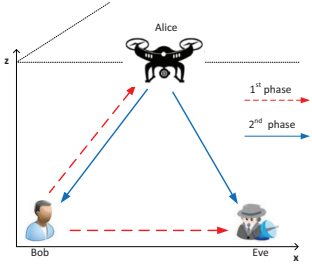}}
\caption{AN-aided secure UAV communications system model.}
\label{fig1}

\end{figure}

We consider a UAV-based wireless communication system including Alice, Bob, and Eve (see Fig. \ref{fig1}), all equipped with {\em single antenna} operating in {\em half-duplex} mode. We assume that Bob and Eve have {\em fixed locations} with three-dimensional (3D) Cartesian coordinates $\omega_b=(x_b,y_b,0)$ and $\omega_e=(x_e,y_e,0)$, respectively, which are known to Alice.   
We define Alice's {\em flying duration} as $T$ seconds and her {\em constant flying height} 
as $H$ meters. Note that UAV's altitude can be time-varying in general, nonetheless, from a practical perspective, this parameter here is considered fixed and chosen to be the minimum allowable height for the purpose of I) avoiding energy consumption of UAV when rising/falling during the mission II) keeping  the UAV away from any possible collision with environmental buildings and surroundings. This leads to establishing likely line-of-sight (LoS) links with ground terminals. Furthermore, Alice's {\em 3D location} at time $t\in [0,T]$ is given by $\omega_a(t)=(x_a(t), y_a(t), H)\in \mathbb{R}^{1\times 3}$. We adopt the widely used
LoS-dominated AG links assumption (see previous research works and references \cite{zhang2019securing, Hongliang2018Sec, Li2019Coo}), since it offers a good approximation to AG links, particularly for rural areas \cite{Lin2018UAV}. Assuming that the Doppler effect arising from UAV’s mobility is perfectly compensated at the terrestrial node $g\in \{b,e\}$, the channel power gain between Alice and any ground terminal is
$\tilde{h}_{ag}(t) = \frac{\beta_0}{\|\omega_a(t)-\omega_g \|^2}$, where $\beta_0 = (\frac{C}{4\pi f_c})^2$ denotes the reference channel power gain for unit distance, wherein $C$ is the speed of light, $f_c$ is the carrier frequency. Further, we have $\omega_g = (x_g,y_g,0) \in \mathbb{R}^{1\times3}$, and $\|\cdot\|$ denotes the $L_2$-norm operator.
 
Descretizing Alice's flying time horizon into $N$ equal-duration time slots, i.e., $\delta_t\treq\frac{T}{N}$, with $\delta_t$ representing slot length, Alice's {\em 3D continuous trajectory} over $T$, i.e., $\omega(t)$ can be approximated by the sequence $\{\omega_a[n]\}^{N}_{n=1}$, such that $\omega_a[n]=(x_a[n],y_a[n],H),~\forall n\in \mathbb{N}$, where $\mathbb{N}=\{1,2,\cdots,N\}$. Note that by choosing $\delta_t$ to be sufficiently small, we can consider that Alice's 3D location is approximately unchanged within each time slot, satisfying invariant channel condition. Besides, we define her {\em predetermined initial and final locations} as $ \omega_{ai}\in \mathbb{R}^{1\times 3}$ and $\omega_{af}\in \mathbb{R}^{1\times 3}$, respectively. 
Let $\bar{d}_\delta \treq \bar{V}\delta_t$, wherein $\bar{V}$ represents Alice's fixed speed, and $\bar{d}_\delta$ denotes Alice's maximum displacement per time slot, yields the  mobility constraints as
\begin{align}
&\mathrm{C1}:\quad   \|\omega_a[1]-\omega_{ai} \| \leq  \bar{d}_\delta,\\
&\mathrm{C2}:\quad    \|\omega_a[n+1]-\omega_a[n] \| \leq  \bar{d}_\delta,~~\mathrm{for}~n\in \mathbb{N} \setminus N \\
&\mathrm{C3}:\quad    \|\omega_{af}-\omega_a[N] \| \leq  \bar{d}_\delta,
\end{align} 

\section{Transmission Protocol}\label{TxProtocol}
{\vspace{-1mm}}
We present a UAV-assisted AN injection based two-phase transmission protocol. We apply frequency division duplexing (FDD) such that both phases of transmission are conducted simultaneously during the coherence time of the wireless channel with {\em equally shared bandwidth}. Also, we assume that the air-ground channel parameters follow reciprocity.

In the {\em first phase}, at time slot $n$, Bob broadcasts an {\em unknown noise-like signal} $z[n]$ (known as pseudo random AN \cite{He2017AN}) of unit-power, i.e., $\E\{\|z[n]\|^2\}=1$, where $\E\{\cdot\}$ is the expectation operator. This signal $z[n]$ is weighted with $\sqrt{P_b[n]}$ to result in transmitted power $P_b[n]$ per transmission. Thus, Alice and Eve respectively receive 
\vspace*{-2mm}
\begin{eqnarray}\label{y1}
y^{(1)}_{a}[n] = \sqrt{P_b[n] \tilde{h}_{ba}[n]}z[n]+\nu_a[n],~~~\forall n\in \mathbb{N}\\
y^{(1)}_{e}[n] = \sqrt{P_b[n]}{g}_{be}[n]z[n]+\nu_e[n],~~~\forall n\in \mathbb{N}
\end{eqnarray}
where the ground-to-ground (G2G) channel power gain follow exponential distribution \cite{Mamaghani2018sec}, i.e., $\|{g}_{be}[n]\|^2\sim \mathrm{Exp}(\lambda_{be})$ with the scale parameter $\lambda_{be} = \frac{\beta_0}{\|\omega_b-\omega_g \|^\eta}$ wherein $\eta>2$, $\nu_a,\nu_e\sim\mathcal{N}(0,N_0)$ are the additive white Gaussian noise (AWGN) with the noise power $N_0$.
Assuming that Bob sends no pilots (training sequences) alongside with $z[n]$, Alice and Eve are unable to obtain the noise-like signal $z[n]$. It should be stressed that reception quality of the G2G link compared to AG link for the first phase is weaker and negligible due to high attenuation, and therefore, this reinforces our assumption that Eve cannot successfully decode $z[n]$ via, for example, joint processing of received signals in two phases.

In the {\em second phase}, at the same time slot, Alice generates 
\vspace*{-4mm}
\begin{align}
 x_a[n] = \sqrt{\alpha[n]}s[n] + \sqrt{1-\alpha[n]}\frac{y^{(1)}_{a}[n]}{\sqrt{\E\{\|y^{(1)}_{a}[n]\|^2\}}},
 \end{align}
where $s[n]$ is the {\em unit-power information signal} and $\alpha[n]$ is the {\em power allocation factor} satisfying
\vspace*{-1mm}
{
\begin{align}
&\mathrm{C4}:\quad    0 \leq \alpha[n] \leq 1,~~~\forall n\in \mathbb{N} 
\end{align}}
Thereafter, Alice forwards information-bearing signal $x_a[n]$ with transmit power $P_a[n]$ over the channel $\tilde{h}_{ag}[n]$, yielding 
{
\begin{align}\label{yk}
\tilde{y}^{(2)}_g[n]&=\stackrel{}{\underset{\text{Information-bearing signal}}{\underbrace{{\sqrt{\alpha[n]P_a[n]\tilde{h}_{ag}[n]}s[n] }}}} \nonumber\\&+\stackrel{}{\underset{\text{AN interference}}{\underbrace{{\sqrt{\frac{(1-\alpha[n])P_a[n]P_b[n] \tilde{h}_{ba}[n]\tilde{h}_{ag}[n]}{P_b[n] \tilde{h}_{ba}[n]+N_0}}z[n]}}}} 
\nonumber\\
&+ \stackrel{}{\underset{\text{Noise}}{\underbrace{ \sqrt {{\frac{(1-\alpha[n])P_a[n]\tilde{h}_{ag}[n]}{P_b[n] \tilde{h}_{ba}[n]+N_0}}} \nu_a[n] +\nu_g[n]}}},~~~\forall n\in \mathbb{N}
\end{align}}
where $\nu_g~\sim~\mathcal{N}(0,N_0)$ is the AWGN. 

In practice, Alice and Bob's transmission powers are subject to the peak powers $\hat{P}_a$ and $\hat{P}_b$ and the average powers $\bar{P}_a$ and $\bar{P}_b$ constraints for all time slots $\forall n\in \mathbb{N}$ respectively as
{\small
\begin{align}
\mathrm{C5}&:~\frac{1}{N}\sum_{n=1}^{N}P_a[n] \leq  \bar{P}_a,~~\mathrm{C6}:~0 \leq P_a[n] \leq \hat{P}_a,\\
\mathrm{C7}&:~\frac{1}{N}\sum_{n=1}^{N}P_b[n] \leq  \bar{P}_b,~~\mathrm{C8}:~0 \leq P_b[n] \leq \hat{P}_b,
\end{align}}
To make the constraints $\mathrm{C5}$ and $\mathrm{C7}$ non-trivial, we assume $\bar{P}_a < \hat{P}_a$ and $\bar{P}_b < \hat{P}_b$. Letting $\gamma_0 \triangleq \frac{\beta_0}{N_0}$, for convenience, we define $h_{ag}[n]\triangleq\frac{\tilde{h}_{ag}[n]}{N_0}$, which can be rewritten as 
\begin{align}
h_{ag}[n] =\frac{\gamma_0}{(x_a[n]-x_g)^2+(y_a[n]-y_g)^2+H^2}.
\end{align}
Assuming that Bob knows perfect channel state information (CSI) (similar to \cite{zhang2019securing, Wang2018Joi, Mamaghani2018sec}), $z[n]$, $P_b[n]$, and also other parameters ($N_0$ and $P_a[n]$) are publicly shared by Alice, he is able to remove AN, yielding his signal-to-interference-plus-noise ratio (SINR) as 
\vspace{-2mm}
{
\begin{align} \label{gamma_B}
\gamma_B[n] = \frac{\alpha[n]P_a[n]h_{ab}[n](P_b[n] h_{ab}[n]+1)}{\left[P_b[n]+\left(1-\alpha[n]\right)P_a[n]\right]h_{ab}[n]+1}.
\end{align}}
However, Eve has no information about the weighted $z[n]$, and so the AN term cannot be cancelled out, leading to 
\vspace{-2mm}\begin{align}\label{gamma_E}
\gamma_E[n] = \frac{\alpha[n]P_a[n]h_{ae}[n]}{\left(1-\alpha[n]\right)P_a[n]h_{ae}[n]+1}.
\end{align}

According to \cite{Milad2019UAVAccess}, the {\em instantaneous secrecy capacity} of the proposed UAV-aided communications in bps/Hz is defined as 
\vspace{-2.5mm}
\begin{align}
    {R}_{sec}[n] \triangleq \frac{1}{2}\left[C_B[n] - C_E[n]\right]^+, ~~~\forall n\in \mathbb{N}
\end{align}
with $C_B[n] \triangleq \log_2(1+\gamma_B[n])$~and~$C_E[n] \triangleq \log_2(1+\gamma_E[n])$,
where $[x]^+\treq\max\{x,0\}$, and fraction $\frac{1}{2}$ is due to half-duplexing. Defining $\pmb{\omega_a}\triangleq\{\omega_a[n]\}_{n=1}^N$,~~  $\pmb{\alpha}\triangleq\{\alpha[n]\}_{n=1}^N$, $\mathbf{P_a} \triangleq\{P_a[n]\}_{n=1}^N$, and $\mathbf{P_b} \triangleq\{P_b[n]\}_{n=1}^N$
we focus on maximizing the {\em average secrecy rate} (ASR) of the proposed scheme over $N$ time slots formulated as
\vspace{-4mm}
\begin{align}\label{opt_prob}
(\mathrm{P}):~~~& \stackrel{}{\underset{\mathbf{P_a}, \mathbf{P_b}, \pmb{\alpha}, \pmb{\omega_a}}{\mathrm{max}}~~\bar{{R}}_{sec} = \frac{1}{N} \sum_{n=1}^{N}  {R}_{sec}[n]} \nonumber\\
&~~~~~\text{s.t.}~~~~~\mathrm{C1-C8},  
\end{align}
The optimization problem $(\mathrm{P})$ is challenging to solve due to non-concave as well as non-differentiable objective function owing to the operator $[\cdot]^+$. At the optimal point, $[\cdot]^+$ can be removed from ${R}_{sec}[n]$ (otherwise, by setting $\mathbf{P_a}=\mathbf{0}$ one can obtain zero ASR), making at least differentiable objective function. Nonetheless, it is still unsolvable due to being a non-convex problem with respect to the optimization variables.

 

\section{Proposed Iterative Algorithm}\label{PropAlgorithm}
To solve problem $(\mathrm{P})$, we propose a computationally tractable iterative algorithm based on BCD and SCA approaches such that problem $(\mathrm{P})$ is split into four subproblems wherein at each time, we optimize one block of variables using convex optimization while keeping the others unchanged in an alternative manner until convergence. To detail the procedure, we omit the constant factor of $\frac{1}{2N\ln 2}$ from the objective function of \eqref{opt_prob} for the sake of convenience.

\subsection{Optimizing Alice's Transmit Power \texorpdfstring{$\mathbf{P_a}$}{Lg} }
For Alice's transmit  power optimization, the problem $(\mathrm{P})$ can be reformulated as 
\begin{align}\label{opt_srcpow}
(\mathrm{P1}):& \stackrel{}{\underset{\mathbf{P_a}}{\mathrm{max}}\sum_{n=1}^{N}\ln\left(1+\frac{a_n P_a[n]}{P_a[n]+b_n}\right) - \ln\left(1+\frac{c_n P_a[n]}{P_a[n]+d_n}\right)} \nonumber\\
&~~\text{s.t.}~~~~~ \mathrm{C5}~\mathrm{and}~\mathrm{C6},
\end{align}
where $a_n = \frac{\alpha[n]}{1-\alpha[n]}\left(1+P_b[n] h_{ab}[n]\right)$,~$b_n=\frac{P_b[n]h_{ba}[n]+1}{(1-\alpha[n])h_{ab}[n]}$,~$c_n=\frac{\alpha[n]}{1-\alpha[n]}$,~$\mathrm{and}~d_n  =\frac{1}{(1-\alpha[n])h_{ae}[n]}$. Using the following lemma, we see that the objective function of the problem $(\mathrm{P1})$ is in concave-minus-concave form.

\begin{lemma}\label{pa_cvx}
Define $f(x)\triangleq\ln(1+\frac{ax+b}{cx+d}), x\geq 0$ with positive constant values $a, b, c, d > 0$ and subject to the condition $ad\geq bc$. $f(x)$ is non-decreasing and has  the first-order derivative given by
\begin{align}
   f'(x) \treq \diffp[1]{f(x)}x = \frac{ad-bc}{(cx+d)(x(a+c)+b+d)},
\end{align}
By taking its second derivative, it can be obtained that the function $f(x)$ is concave as per the second-order condition law which states that $f(x)$ is concave iff $\diffp[2]{f(x)}x\leq 0$. Further, since the first order Taylor approximation at a given point $x_0$ of a concave function provides a global overestimator at that point \cite{cvx_boyd}, the inequality below holds 
 \begin{align}
    f(x) &\leq f(x_0)+f'(x_0)(x-x_0)\nonumber\\
    &\leq \ln(1+\frac{ax_0+b}{cx_0+d}) + \frac{(ad-bc)(x-x_0)}{(cx_0+d)(b+d+(a+c)x_0)}.
\end{align}
\end{lemma}
Therefore, we approximate the problem $(\mathrm{P1})$ with the following convex alternative as {\vspace{-1mm}}
 \begin{align}\label{opt_srcpow_cvx}
(\mathrm{P2}):& \stackrel{}{\underset{\mathbf{P_a}}{\mathrm{max}}~~\sum_{n=1}^{N}\ln\left(1+\frac{a_n P_a[n]}{P_a[n]+b_n}\right) - A_n P_a[n]   } \nonumber\\
&~~\text{s.t.}~~~~~ \mathrm{C5}~\mathrm{and}~\mathrm{C6},
\end{align}
where $
A_n = \frac{c_nd_n}{(P^{(k)}_a[n]+d_n)(d_n+(c_n+1)P^{(k)}_a[n])},$
with
the feasible points $\mathbf{P}^{(k)}_\mathbf{a}\triangleq \{P^{(k)}_a[n]\}_{n=1}^N$. We note that the objective value of $(\mathrm{P2})$ is lower-bounded by that of $(\mathrm{P1})$ at $\mathbf{P}^{(k)}_\mathbf{a}$. Since $(\mathrm{P2})$ is convex, thus, it can be solved optimally using CVX.

\subsection{Optimizing Bob's Transmit Power \texorpdfstring{$\mathbf{P_b}$}{Lg}}
By keeping the other variables unchanged, we optimize Bob's transmit power via reformulating the problem $(\mathrm{P})$, ignoring the constant second term of summation, as
\begin{align}\label{opt_despow}
(\mathrm{P3}):& \stackrel{}{\underset{\mathbf{P_b}}{\mathrm{max}}~~\sum_{n=1}^{N}\ln\left(1+\frac{k_0 P_b[n]+k_1}{k_2P_b[n]+k_3}\right)} \nonumber\\
&~~\text{s.t.}~~~~~ \mathrm{C7}~\mathrm{and}~\mathrm{C8},
\end{align}
where $k_0 = \alpha[n]P_a[n]h^2_{ab}[n]$, $k_1 = \alpha[n]P_a[n] h_{ab}[n]$, $k_2=h_{ab}[n]$, and $k_3=(1-\alpha[n])P_a[n]h_{ab}[n]+1$. Following Lemma \ref{pa_cvx}, $(\mathrm{P}3)$ is a convex problem, since it has concave objective function with convex constraints. Thus, it can be solved optimally via Lagrangian method as $(\mathrm{P}3)$ satisfies the Slater's condition and strong duality holds. As such, by temporarily dropping $\mathrm{C8}$ and  also letting $\tilde{\mathbf{P}}_\mathbf{b}$ and $(\tilde{\mathbf{P}}_\mathbf{b}, \lambda)$ be any primal and dual optimal points, only satisfying Karush-Kuhn-Tucker (KKT) conditions results in zero duality gap.
As such, the Lagrangian of $(\mathrm{P}3)$ is given by
\begin{align}
\hspace{-3mm}\mathcal{L}\left(\mathbf{P}_b,\lambda\right) \hspace{-1mm}=\hspace{-1mm} -\hspace{-1mm}\sum_{n=1}^{N}\ln\hspace{-1mm}\left(\hspace{-1mm}1\hspace{-1mm}+\hspace{-1mm}\frac{k_0 P_b[n]\hspace{-1mm}+\hspace{-1mm}k_1}{k_2P_b[n]\hspace{-1mm}+\hspace{-1mm}k_3}\hspace{-1mm}\right) \hspace{-1mm}+\hspace{-1mm} \lambda (\sum_{n=1}^{N}\hspace{-1mm}P_b[n]\hspace{-1mm}-\hspace{-1mm}P^{tot}_b),
\end{align}
where $\lambda \geq 0$ denotes the {\em Lagrange multiplier} associated with the inequality $\mathrm{C7}$, and $P^{tot}_b\treq N\bar{P}_b$. Then, maximizing the Lagrangian dual function defined as $g(\lambda) \treq~\stackrel{}{\underset{\mathbf{P_b}}{\inf}}~\{\mathcal{L}\left(\mathbf{P}_b,\lambda\right)\},$
one can attain the optimality condition for $\forall n\in\mathbb{N}$ as
\begin{align}
\frac{k_0k_3-k_1k_2}{(k_3+k_2P_b[n])(P_b[n](k_0+k+2))+k_1+k_3} \hspace{-1mm}-\hspace{-1mm}\lambda = 0,
\end{align}
Solving the above equation with respect to $P_b[n]$ and also taking into account constraint $\mathrm{C8}$, leads to the closed-form analytical solution for  $(\mathrm{P}3)$ as 
\begin{align}\label{bob_optpow}
P^{\star}_b[n]= \min\left\{\hat{P}_b, \frac{\sqrt{a^2_1-4a_0a_2}-a_1}{2a_2}\right\},~\forall n\in \mathbb{N}
\end{align}
where $a_2=k_2(k_0+k_2)$, $a_1=k_1k_2+2k_2k_3+k_0k_3$, and $a_0 = k_3(k_1+k_3)- \frac{k_0k_3-k_1k_2}{\lambda}$, wherein the non-negative  Lagrange factor $\lambda$ can be obtained by applying a simple bisection search such that Bob's power budget constraint; i.e., $\sum_{n=1}^{N}P^\star_b[n] \leq P^{tot}_b$ is satisfied.


\subsection{Optimizing power allocation factor
\texorpdfstring{$\pmb{\alpha}$}{Lg} }
To optimize the third block of variables $\pmb{\alpha}$, we recast the problem $(\mathrm{P})$ equivalently as
\begin{align}\label{opt_paf}
(\mathrm{P4}):& \stackrel{}{\underset{\pmb{\alpha}}{\mathrm{max}}~~\sum_{n=1}^{N}\ln\left(\Psi(\alpha[n])\right)} \nonumber\\ 
&~~\text{s.t.}~~~~~ \mathrm{{C4}}: \quad 0\leq \alpha[n] \leq 1,~~~\forall n\in \mathbb{N}
\end{align}
where the function $\Psi(x)$ is defined as
\begin{align}
    \Psi(x) = \frac{[(1-x)\gamma_3+1][(x\gamma_2+1)\gamma_1+\gamma_2+1]}{(\gamma_3+1)[(1-x)\gamma_1+\gamma_2+1]},
\end{align}
wherein $\gamma_1 \treq P_a[n] h_{ab}[n]$, $\gamma_2 \treq P_b[n] h_{ab}[n]$,~$\gamma_3 \treq P_a[n] h_{ae}[n]$. We note that $\sum([\cdot]^+)$ is a non-decreasing affine function and  $\ln(\Psi(x))$ is a monotonically increasing function with respect to $\Psi(x)$. Therefore, $(\mathrm{P4})$ can be solved by maximizing every single term of the summation. Accordingly, optimized power allocation factor $\alpha^\star[n]$ for $\forall n\in \mathbb{N}$ can be calculated as
\begin{align}\label{alpha_opt}
\alpha^\star[n]& = \frac{1\hspace{-0.5mm}+\hspace{-0.5mm}\gamma_1\hspace{-0.5mm}+\hspace{-0.5mm}\gamma_2}{\gamma_1}\hspace{-0.5mm}-\hspace{-0.5mm}\frac{\sqrt{(\gamma_2\gamma_3\hspace{-0.5mm}-\hspace{-0.5mm}\gamma_1)(1\hspace{-0.5mm}+\hspace{-0.5mm}\gamma_2)(1\hspace{-0.5mm}+\hspace{-0.5mm}\gamma_1\hspace{-0.5mm}+\hspace{-0.5mm}\gamma_2)\gamma_2\gamma_3}}{\gamma_1\gamma_2\gamma_3}\nonumber\\
&\stackrel{(a)}{\approx} 1-\left[\sqrt{\frac{P_b[n]}{P_a[n]}\left(1+\frac{P_b[n]}{P_a[n]}\right)}-\frac{P_b[n]}{P_a[n]}\right],
\end{align}
where $(a)$ follows from high SNR approximation, i.e., $\gamma_i\gg 1$ for $i\in\{1, 2, 3\}$.
\begin{proof}
We claim that $\Psi (x)$ is a quasi-concave function of $x$ in the feasible set $0 \leq x \leq 1$. Indeed, since the second derivative $\diffp[2]{\Psi(x)}x$ is quite sophisticated to argue about the convexity of the function, our approach is to use the properties of the first derivative of the function $\Psi(x)$ with respect to $x$. After tedious calculation, we can construe that for practical values of $\gamma_1,\gamma_2, \gamma_3\geq 1$, we have $\diff[1]{\Psi}x[0]=  \frac{\gamma_1(\gamma_2+1)}{1+\gamma_1+\gamma_2}-\frac{\gamma_3}{\gamma_3+1} > 0$, and 
$\diff[1]{\Psi}x[1]= \frac{\gamma_1+\frac{\gamma^2_1}{\gamma_2+1}-\gamma_3(1+\gamma_1)}{\gamma_3+1}
 < 0 $. Besides, it can be readily seen, via solving $\diffp[1]{\Psi}x=0$, that $\Psi(x)$ has only one extremum, say $x^\star$, in the domain $[0, 1]$,
leading to the fact that the function $\Psi(x)$ is increasing in $[0, x^\star)$, and tends to be a decreasing function in  $(x^\star, 1]$, and thus, the proof is done. 
\end{proof}

\subsection{Optimizing UAV-Alice's trajectory 
\texorpdfstring{$\pmb{\omega_a}$}{Lg} }
The corresponding subproblem to optimize Alice's trajectory is approximately reformulated, by taking the slack variables $\mathbf{s}=\{s[n]\}_{n=1}^{N}$ and $  \mathbf{v}=\{v[n]\}_{n=1}^{N}$, as 
 \begin{align}\label{Traj_opt}
(\mathrm{P5}):& \stackrel{}{\underset{\pmb{\omega_a}, \mathbf{s}, \mathbf{v}}{\mathrm{max}}~~\sum_{n=1}^{N}\ln\left(c_0+\frac{c_1}{s[n]}\right) - \ln\left(1+\frac{c_2}{c_3+v[n]}\right)} \nonumber\\
&~~\text{s.t.}~~~~~ \mathrm{C1-C3,}\nonumber\\
&\mathrm{C9}:\quad   \|\omega_a[n]-\omega_b \|^2 \leq s[n],~~~\forall n\in \mathbb{N}\nonumber\\
&\mathrm{C10}:\quad  \|\omega_a[n]-\omega_e \|^2 \geq v[n],~~~\forall n\in \mathbb{N}
\end{align} 
where $c_0 = \frac{\alpha[n]P_a[n]}{P_b[n]+P_a(1-\alpha[n])}$,~$c_1 = \frac{\alpha[n]P_a[n]P_b[n]\gamma_0}{P_b[n]+P_a(1-\alpha[n])}$, $c_2 = \alpha[n]P_a[n]\gamma_0$,~$c_3=(1-\alpha[n])P_a[n]\gamma_0$. Note that at the optimal point, constraints $\mathrm{C9}$ and $\mathrm{C10}$ must hold with equality, otherwise, by varying $s[n]$ ($v[n]$) the  value of the objective function in $(\mathrm{P5})$ increases, and this, of course, violates the optimality.  Before solving $(\mathrm{P5})$, we mention some fruitful lemmas below.
\begin{lemma}\label{trj_cvx}
Define the function $g(x) \triangleq \ln \left(a+b x^{-c}\right), x\geq 0$ with the first and second derivatives given respectively by
\begin{align}
\frac{\partial g(x)}{\partial x} \hspace{-0.5mm}=\hspace{-0.5mm}  -\frac{b c}{a x^{c+1}\hspace{-0.5mm}+\hspace{-0.5mm}b x},~ \frac{\partial^2 g(x)}{\partial x^2} \hspace{-0.5mm}=\hspace{-0.5mm}  \frac{b c \left(a (c\hspace{-0.5mm}+\hspace{-0.5mm}1) x^c\hspace{-0.5mm}+\hspace{-0.5mm}b\right)}{x^2 \left(a x^c\hspace{-0.5mm}+\hspace{-0.5mm}b\right)^2},
\end{align}
 where  the constants $a, b, c$ hold non-negative values. Having the non-negative second order derivative, the function $g(x)$ is convex, and since according to the first-order convexity condition which states that for a convex function the first-order Taylor approximation is a global affine underestimator of that function and vice versa \cite{cvx_boyd}, therefore, $g(x)$ has the following global lower bound at the given point $x_0$ as
\begin{align}
     g(x) \geq \ln \left(a+\frac{b}{x^{c}_0} \right) -\frac{b c}{x_0\left(a x^{c}_0+b \right)}(x-x_0),
\end{align}
\end{lemma}

\begin{lemma}\label{over_estimator}
	Let $\mathbf{x}$ be a vector of variables $\{x_i\}_{i=1}^{N}$ and  $\mathbf{a} \in \mathbb{R}^{N\times1}$  be a constant vector . The function of negative norm-squared of this two vectors; $h(\mathbf{x}) = \|\mathbf{x}-\mathbf{a}\|^2$, which obviously is a convex function with respect to the vector $\mathbf{x}$, has a global concave lower-bound, according to first-order convexity condition in Lemma \ref{trj_cvx}, given by
	 \begin{align}
	\|\mathbf{x}-\mathbf{a}\|^2 \geq -\|\mathbf{x_0}\|^2 + 2\left(\mathbf{x_0} - \mathbf{a}\right)^\dagger\mathbf{x} + \|\mathbf{a}\|^2,
	\end{align}
	wherein $(\cdot)^\dagger$ represents the transpose operator.
	\end{lemma}
Based on Lemma \ref{trj_cvx}, we see that the objective function of problem $(\mathrm{P5})$ has the convex-minus-convex form, and therefore, replacing the first convex term of which with the corresponding global concave approximation, similarly for the convex function $\|\omega_a[n]-\omega_e \|^2$ in constraint $\mathrm{C10}$ using Lemma \ref{over_estimator}, yields an approximated  lower-bound convex problem under the given feasible points
$\pmb{\omega}^{(k)}_\mathbf{a}\triangleq \{{\omega^{(k)}_a[n]}\}_{n=1}^N$ 
$\mathbf{s}^{(k)}\triangleq \{s^{(k)}[n]\}_{n=1}^N$ as
\begin{align}\label{Traj_opt_approx}
(\mathrm{P6}):& \stackrel{}{\underset{\pmb{\omega_a}, \mathbf{s}, \mathbf{v}}{\mathrm{max}}~~\sum_{n=1}^{N}B_ns[n] - \ln\left(1+\frac{c_2}{c_3+v[n]}\right)} \nonumber\\
&\text{s.t.}~~~~~ \mathrm{C1-C3, C9},\nonumber\\ {\vspace{-1mm}}
&\hspace{-5mm}\widetilde{C10}:\quad v[n]  - 2\left({\omega^{(k)}_a}[n] - {\omega_e}\right)^\dagger{\omega_a[n]} +c_4 \leq 0,
\end{align}
where $B_n = -\frac{c_1}{s^{(k)}[n]\left(c_0+c_1 \right)}$, $c_4 \treq \|{\omega^{(k)}_a}[n]\|^2 - \|{\omega_e}\|^2$. 
Being a convex problem, $(\mathrm{P6})$ can be solved efficiently using CVX. 
\begin{figure}
	\centering
\resizebox{\columnwidth}{!}{%
 \removelatexerror
  \begin{algorithm}[H] \label{algorithm1}
  \caption{Proposed iterative algorithm}
  1:~\textbf{Initialize}~
  {feasible points $\pmb{\omega}^{(0)}_\mathbf{a}$, $\pmb{\alpha}^{(0)}$, $\mathbf{P}^{(0)}_\mathbf{a}$, $\mathbf{P}^{(0)}_\mathbf{b}$, $\mathbf{s}^{(0)}$, $\mathbf{v}^{(0)}$, set convergence tolerance $\epsilon$, and $k=0$;}\\
  2:~\textbf{Repeat:} {~$k\gets k+1;$\\
  3:~Given $\pmb{\omega}^{(k-1)}_\mathbf{a}$, $\pmb{\alpha}^{(k-1)}$, $\mathbf{P}^{(k-1)}_\mathbf{b}$, and $\mathbf{P}^{(k-1)}_\mathbf{a}$ solve $(\mathrm{P2})$ updating  $\mathbf{P}^{(k)}_\mathbf{a}$;\\
  4:~Given $\pmb{\omega}^{(k-1)}_\mathbf{a}$, $\pmb{\alpha}^{(k-1)}$, $\mathbf{P}^{(k)}_\mathbf{a}$, and $\mathbf{P}^{(k-1)}_\mathbf{b}$ solve $(\mathrm{P3})$ using \eqref{bob_optpow} updating  $\mathbf{P}^{(k)}_\mathbf{b}$;\\
  5:~Given $\mathbf{P}^{(k)}_\mathbf{a}$, $\mathbf{P}^{(k)}_\mathbf{b}$ $\pmb{\omega}^{(k-1)}_\mathbf{a}$, update  $\pmb{\alpha}^{(k)}$ using \eqref{alpha_opt};\\
  6:~Given $\pmb{\omega_a}^{(k-1)}$, $\mathbf{s}^{(k-1)}, \mathbf{v}^{(k-1)}$, $\pmb{\alpha}^{(k)}$, $\mathbf{P}^{(k)}_\mathbf{a}$, and $\mathbf{P}^{(k)}_\mathbf{b}$, solve $(\mathrm{P5})$, then update  $\pmb{\omega}^{(k)}_\mathbf{a}$, $\mathbf{s}^{(k)}$, and $\mathbf{v}^{(k)}$;}\\
  7:~\textbf{Until} {the fractional increase of the objective function is below the threshold $\epsilon$;}
  \end{algorithm}}
\end{figure}

\vspace{-9mm}
\subsection{Overall Algorithm}
The proposed iterative algorithm is summarized in Algorithm \ref{algorithm1}, whose complexity is dominated by the complexity of each sub-problem. As such, sub-problem (P1) is solved by SCA method and its complexity mainly relies on the number of variables and constraints. Since there are $N+1$ constraints in (P2), the number of iterations required for SCA is $\mathcal{O}(\sqrt{N+1}\log_2(\frac{1}{\varepsilon_1}))$, where $\varepsilon_1$ is the accuracy of SCA method for solving (P2). Besides, the complexity of solving (P2) at each iteration is $\mathcal{O}(N^2(N+1))$. Consequently,  the complexity of solving (P2) can be approximately represented as $\mathcal{O}(N^{3.5}\log_2(\frac{1}{\varepsilon_1}))$. We solved (P3) semi-analytically based on Lagrangian method and using a bisection search whose complexity is $\log_2(\frac{\varepsilon_0}{\varepsilon_3})$, where $\varepsilon_0$ and $\varepsilon_3$ represent initial bracket size of $[\lambda_{min}, \lambda_{max}]$, and the required tolerance, respectively. Further, (P4) is solved optimally at one iteration. Similar to that of (P1), the complexity of (P5) is approximately $\mathcal{O}((4N)^2(3N+1)^{1.5}\log_2(\frac{1}{\varepsilon_2}))$. Finally, the overall computational complexity
of the proposed algorithm is approximately 
$\mathcal{O}(M(N^{3.5}\log_2(\frac{1}{\varepsilon_1})+\log_2(\frac{\varepsilon_0}{\varepsilon_3})+48\sqrt{3}N^{3.5}\log_2(\frac{1}{\varepsilon_2})))$, wherein $M$ denotes the number of iterations for Algorithm \ref{algorithm1}, which is inversely proportional to the termination threshold $\epsilon$.

For convergence analysis of Algorithm \ref{algorithm1}, we define the objective values of original problem (P), the subproblems problems (P2), and (P6) at iteration $k$ as 
$\bar{R}_{sec}\left(\mathbf{P}^k_\mathbf{a}, \mathbf{P}^k_\mathbf{b}, \pmb{\alpha}^k, \pmb{\omega}^k_\mathbf{a}\right)$, $\Theta_{lb}\left(\mathbf{P}^k_\mathbf{a},\mathbf{P}^k_\mathbf{b}, \pmb{\alpha}^k, \pmb{\omega}^k_\mathbf{a}\right)$ and $\Xi_{lb}\left(\mathbf{P}^k_\mathbf{a},\mathbf{P}^k_\mathbf{b}, \pmb{\alpha}^k, \pmb{\omega}^k_\mathbf{a}\right)$, respectively. We can write as 
{\small
\begin{align}\label{convergence}
&\bar{R}_{sec}\left(\mathbf{P}^k_\mathbf{a}, \mathbf{P}^k_\mathbf{b},\pmb{\alpha}^k, \pmb{\omega}^k_\mathbf{a}\right)
\stackrel{(a_1)}{\leq}  
\bar{R}_{sec}\left(\mathbf{P}^{k+1}_\mathbf{a}, \mathbf{P}^k_\mathbf{b}, \pmb{\alpha}^k, \pmb{\omega}^k_\mathbf{a}\right)\nonumber\\
&\stackrel{(a_2)}{\leq} \bar{R}_{sec}\left(\mathbf{P}^{k+1}_\mathbf{a}, \mathbf{P}^{k+1}_\mathbf{b}, \pmb{\alpha}^k, \pmb{\omega}^k_\mathbf{a}\right)
\stackrel{(b)}{=} \Theta_{lb}\left(\mathbf{P}^{k+1}_\mathbf{a}, \mathbf{P}^{k+1}_\mathbf{b},\pmb{\alpha}^k, \pmb{\omega}^k_\mathbf{a}\right)\nonumber\\
&\stackrel{(c)}{\leq}
\bar{R}_{sec}\left(\mathbf{P}^{k+1}_\mathbf{a}, \mathbf{P}^{k+1}_\mathbf{b},\pmb{\alpha}^{k+1}, \pmb{\omega}^k_\mathbf{a}\right)\stackrel{(d)}{=} \Xi_{lb}\left(\mathbf{P}^{k+1}_\mathbf{a}, \mathbf{P}^{k+1}_\mathbf{b},\pmb{\alpha}^{k+1}, \pmb{\omega}^k_\mathbf{a}\right)\nonumber
\\
&\stackrel{(e)}{\leq}
\bar{R}_{sec}\left(\mathbf{P}^{k+1}_\mathbf{a}, \mathbf{P}^{k+1}_\mathbf{b},\pmb{\alpha}^{k+1}, \pmb{\omega}^{k+1}_\mathbf{a}\right),
\end{align}}
where $(a_1)$ and $(a_2)$ follows from the definition of the optimal solution to (P1) and (P3), respectively. Besides, the equalities $(b)$ and $(d)$ hold since the first order Taylor approximation is adopted and that the objective functions of (P2) and (P6) share the same value with the original function at the given points, respectively. Finally, $(c)$ and $(e)$ hold since the objective functions of (P2) and (P6)  are tight lower-bounds to that of (P1) and (P5), respectively. The last inequality in \eqref{convergence} indicates that the objective value of (P) is non-decreasing with the iteration index. Since the optimal value of (P) is upper bounded by a finite value, Algorithm \ref{algorithm1} is guaranteed to converge.


\section{Numerical results}\label{numresult}
\begin{table}[t]
	\centering
\caption{Simulation parameters}
\resizebox{0.9\columnwidth}{!}{%
\centering
\begin{tabular}{||c | c || c | c|| c | c||} 
 \hline
 Notation & Value & Notation & Value & Notation & Value \\ [0.5ex] 
 \hline\hline
 $H$              & $100$m       & $\gamma_0$             & 80 dB     & $\lambda$        & $0.5$ \\\hline
 $\omega_b$       & $(0,0,0)$    & $\bar{P}_{ave}$        & 0 dBm     & $ \bar{\alpha}$  & $0.5$  \\\hline 
 $\omega_e$       & $(100,0,0)$  & $\hat{P}_a = \hat{P}_b$& 4$\bar{P}_{ave}$          & $\epsilon$ & $10^{-4}$ \\\hline
 ${\omega_{ai}}$  & $(50,+200,0)$& $\bar{P}_a$            & $\lambda P_{ave}$ &  $\bar{V}$        & $4$ m/s     \\\hline
 ${\omega_{af}}$  & $(50,-200,0)$& $\bar{P}_b$            & $(1-\lambda) P_{ave}$ & $N$                    & 100 \\ \hline
  \hline
\end{tabular}}

\label{table:1}
\end{table}
In this section, we conduct simulations to verify the convergence of Algorithm \ref{algorithm1} and demonstrate the performance of the proposed scheme, labelled as \textit{JTDORA}. Unless otherwise stated, we adopt all parameters in Table \ref{table:1} for simulations. Besides, for initial feasible trajectory of UAV, we adopt the best-effort approach (labeled as \textit{Baseline} trajectory), wherein given Alice's flying duration $T$ seconds, she flies towards Bob with maximum speed in direct path, then hovers right above Bob's location as long as possible, and finally heads directly at maximum speed to the final location. Otherwise, when the hovering phase is impossible due to limited time, while flying at maximum speed to Bob, Alice turns at a midway point to reach the final location by the end of the time. We consider the following benchmarks for performance comparison:
\begin{itemize}
     \item \textit{ANOPC}: AN-based scheme with baseline trajectory, fixed AN injection factor, i.e., $\alpha[n]=\bar{\alpha},~\forall n$, and iteratively optimizing $\mathbf{P}_{\mathbf{a}}$ and  $\mathbf{P}_{\mathbf{b}}$ using $(\mathrm{P2})$ and \eqref{bob_optpow}, respectively. Note that ignoring trajectory design, this scheme is similar to that in \cite{He2017AN}.
     \item \textit{ANTD}: AN-based scheme with trajectory design via iteratively optimizing $\pmb{\omega}_{\mathbf{a}}$ using $(\mathrm{P6})$, fixed transmit powers, $P_a[n]=\bar{P}_a, \forall n$, and $P_b[n]=\bar{P}_b$, $\forall n$, and equal AN injection factor $\alpha[n]=\bar{\alpha}$, $\forall n$.
     \item \textit{ANERA}: AN-based scheme with baseline trajectory and fixed resource allocations, i.e., $P_a[n]=\bar{P}_a, P_b[n]=\bar{P}_b,~\alpha[n]=\bar{\alpha}, \forall n$. This scheme is indeed the initial feasible point $(\pmb{\omega}^{(0)}_\mathbf{a}, \mathbf{P}^{(0)}_\mathbf{a}, \mathbf{P}^{(0)}_\mathbf{b}, \pmb{\alpha}^{(0)})$ used for solving Algorithm \ref{algorithm1}.
      \item \textit{TDPC}: Trajectory optimization and power controlling without AN allocation, i.e., $\alpha[n]=0~ \forall n$. It is quite similar, excluding the duplexing aspect, to that in \cite{zhang2019securing}.  For the fairness of comparisons, we set $\bar{P}_a={P}_{ave}$ to ensure all schemes have an equal total transmission power budget.
 \end{itemize}

\begin{figure}[t]
        \centering
        \includegraphics[width= \columnwidth]{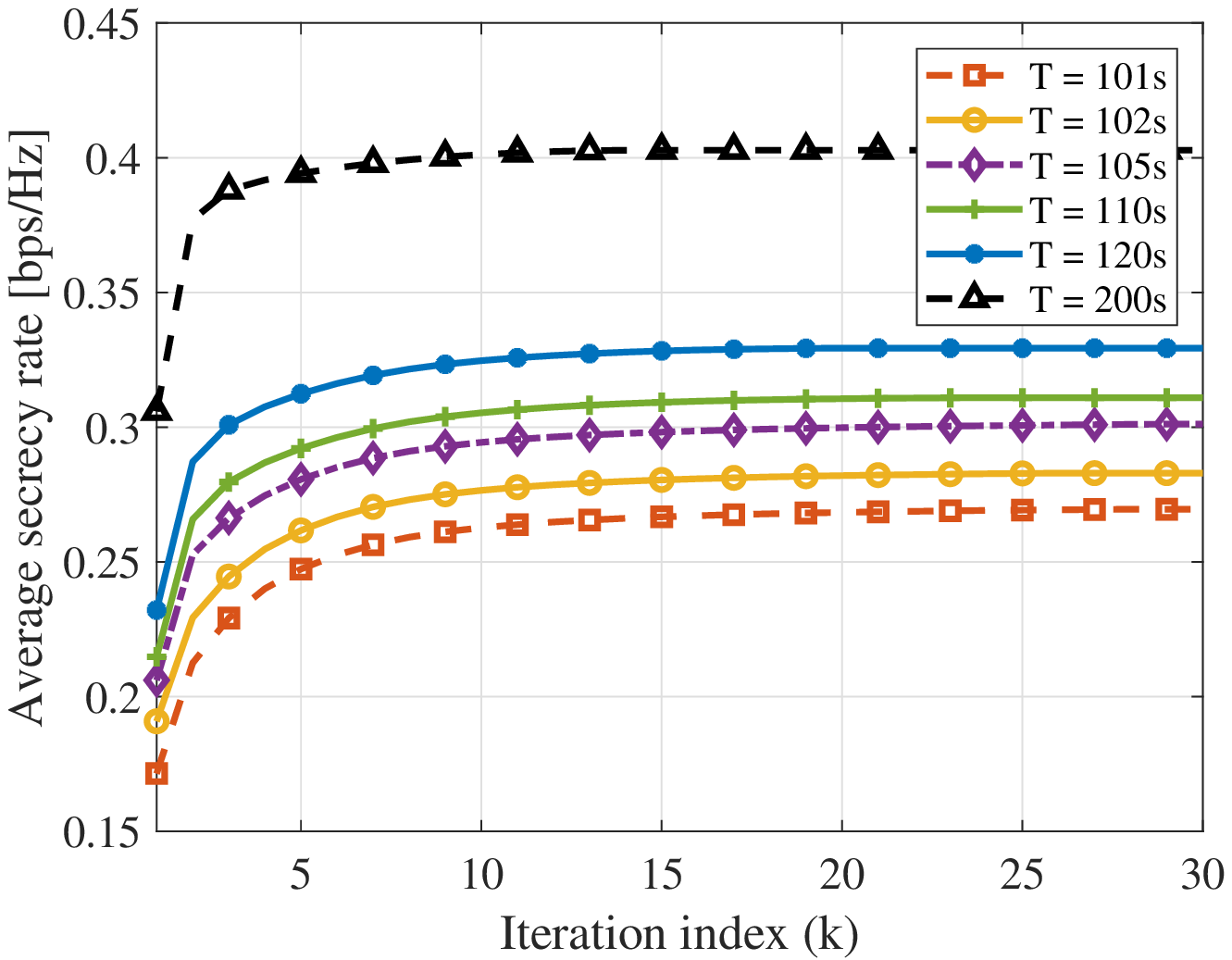}
        \caption{Convergence Verification.}
        \label{sim:fig1}
\end{figure}
 \begin{figure}[t]
        \centering
        \includegraphics[width= \columnwidth]{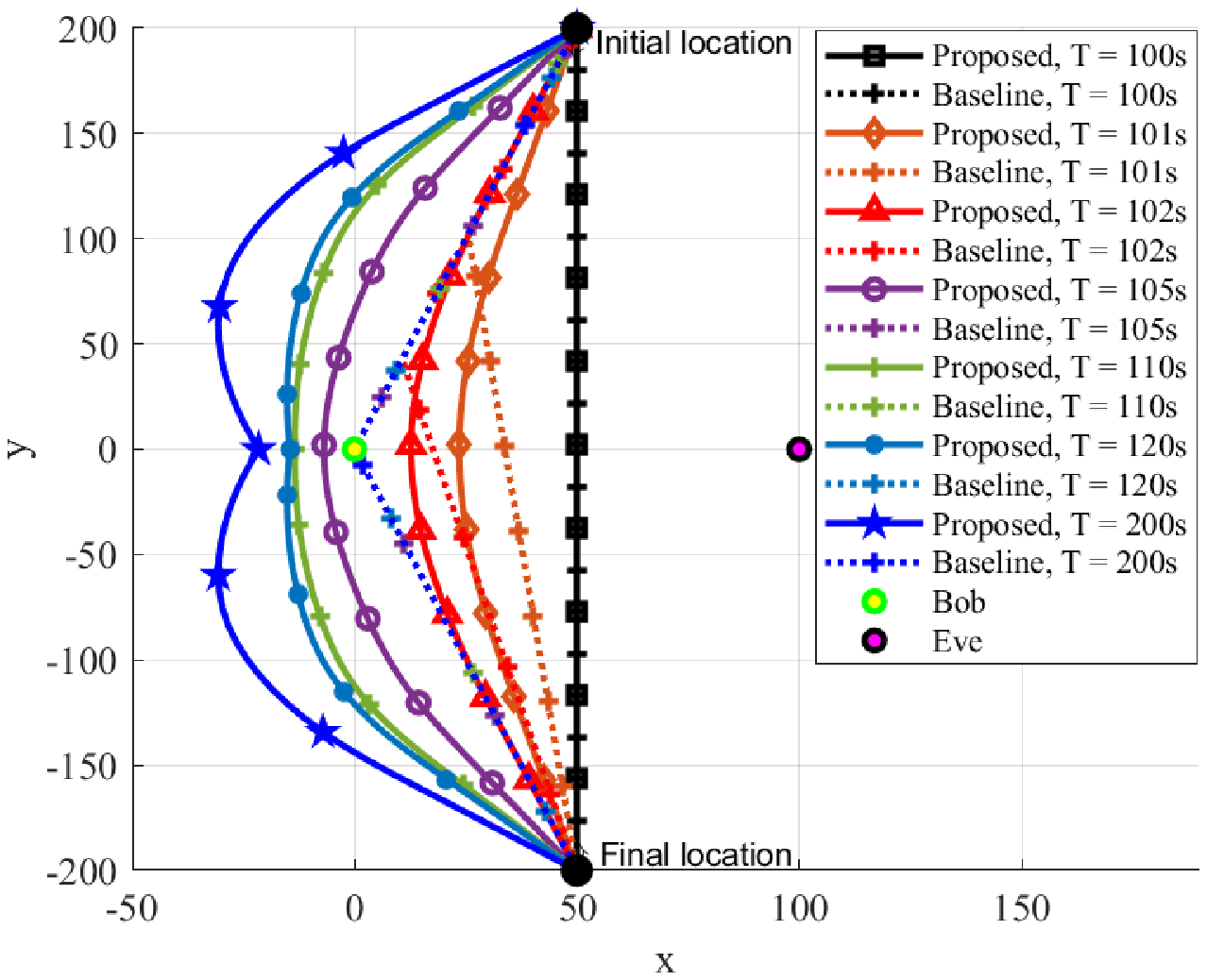}
        \caption{ Alice's designed trajectory.}
        \label{sim:fig2}
\end{figure}
\begin{figure}[t]
    \centering
    \includegraphics[width=\columnwidth]{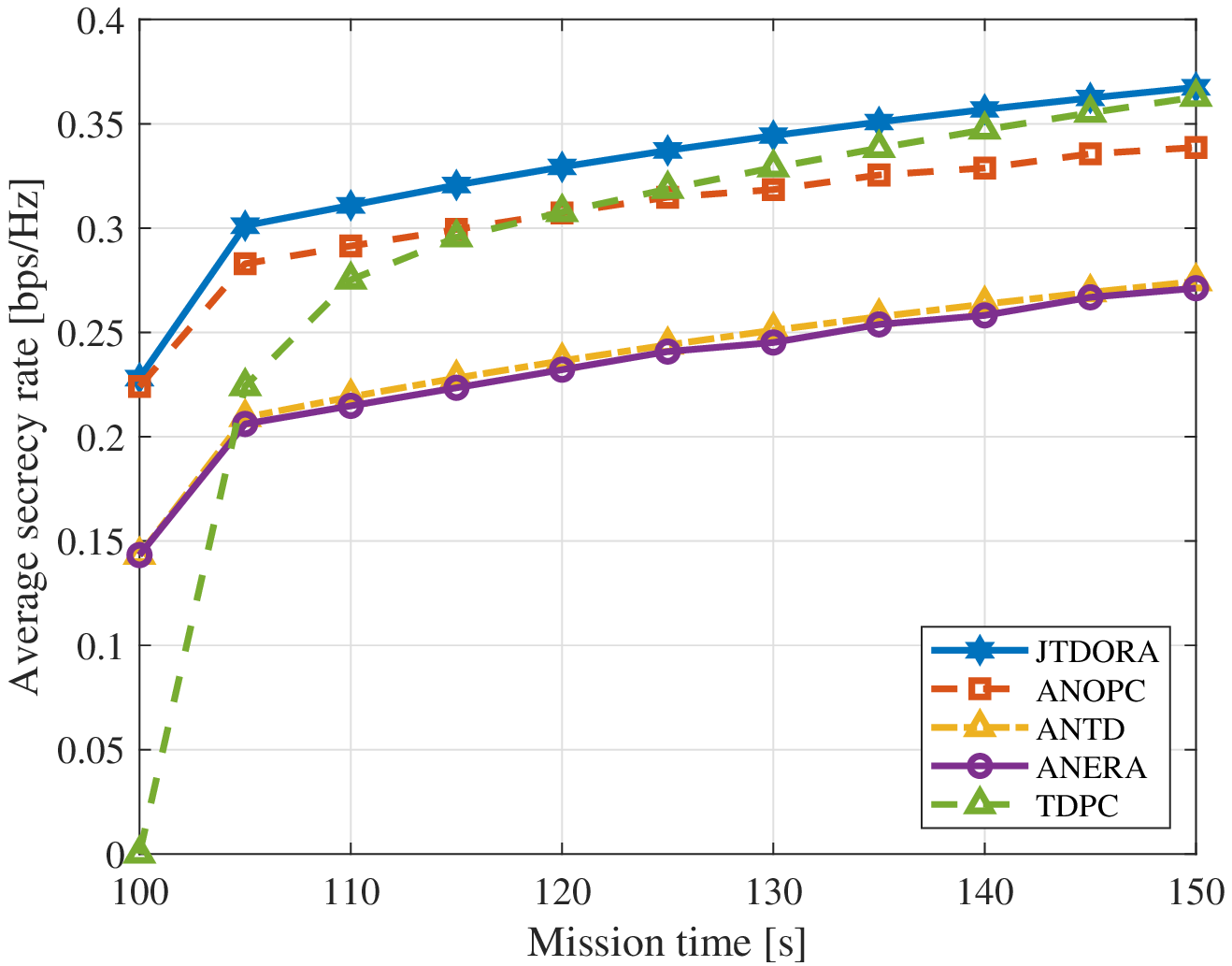}
    \caption{ ASR vs. mission time.}
    \label{sim:fig3}
\end{figure}

\begin{figure}[t]
     \centering
     \includegraphics[width=\columnwidth]{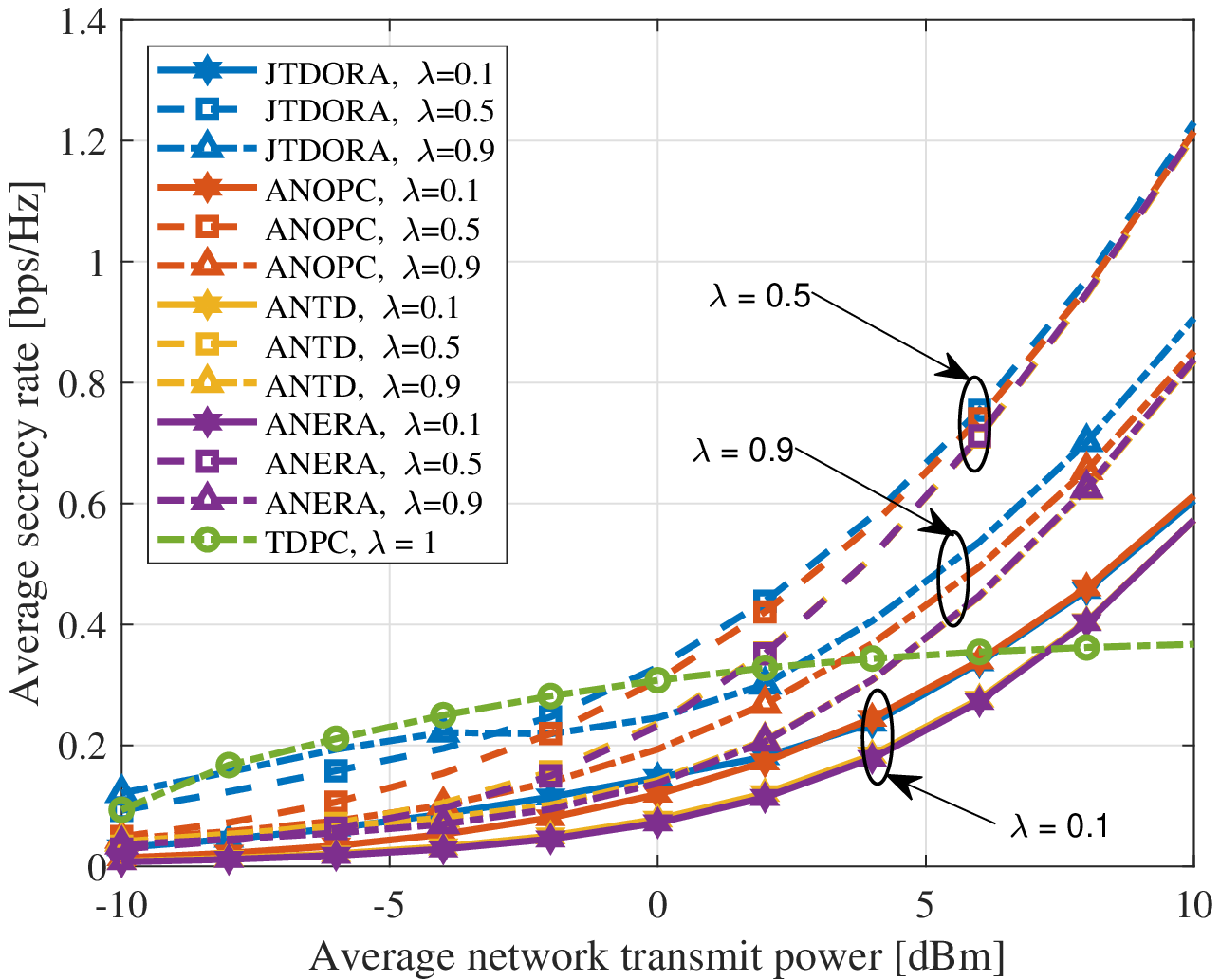}
      \caption{ASR vs. $P_{ave}$.}
        \label{sim:fig5}
\end{figure}

We verify our analysis in Fig. \ref{sim:fig1} by plotting ASR vs iteration indices. We see that the ASR is non-decreasing over the iteration index, and our proposed algorithm converges in very few iterations. Besides, the more the mission time, the higher the achievable ASR, since Alice can spend longer time hovering near Bob. 

In Fig. \ref{sim:fig2} we plot the designed trajectory using Algorithm \ref{algorithm1} vs the baseline trajectory for different flight durations. When $T=100s$, the only feasible path is the direct line from initial to final locations and Alice does not have flexibility in path planning. When $T$ increases, the designed trajectory of Alice gets curved, since this makes Alice fly as close as possible to Bob and as farther as possible from Eve to improve ASR. Note that for baseline trajectory when approximately $T\leq 104s$ holds, Alice can just perform the non-stop midpoint-turn flying approach; otherwise, she can do fly-hover-fly with different hovering durations.
From Fig. \ref{sim:fig2}, we observe, using the \textit{JTDORA}-based trajectory design, the best location for Alice to have a secure communication while hovering, is quite farther from Bob, depending on the mission duration. 

In Fig. \ref{sim:fig3}, we compare the performance of the proposed \textit{JTDORA} with the aforementioned \textit{ANOPC}, \textit{ANTD}, \textit{ANERA}, and \textit{TDPC} benchmark schemes in terms of ASR for different flight durations. For AN-based schemes, we set equal total power allocation, denoted by $\lambda$, between Alice and Bob for information and AN transmission, respectively. We observe that, for the mission times of interest, our proposed \textit{JTDORA} demonstrates the highest ASR amongst all. In particular, when $T\le 104$s, \textit{TDPC} without AN performs the worst, but when $T$ is high (e.g. $T> 104$s), \textit{TDPC} achieves improved ASR, when compared to \textit{ANOPC}, \textit{ANTD}, and \textit{ANERA}, since \textit{TDPC} consumes all transmit power for information transmission. For a large $T$, when UAV is able to find the most convenient location for communication, \textit{TDPC} can highly improve ASR. In contrast, the AN-based schemes spend some fraction of allocated power for AN injection, leaving less power for confidential information transmission. Overall, we conclude that our proposed \textit{JTDORA} scheme is more appropriate for stringent flight mission periods, and $\lambda$ should be chosen dynamically according to the flight duration. 

Fig. \ref{sim:fig5} depicts the impacts of {\em average network power} $P_{ave}$ on ASR for different schemes with different power ratio factors, with $T=120s$. We observe that increasing $P_{ave}$ results in improved ASR for all schemes, whereas for \textit{TDPC}, an ASR ceiling appears in very high average network transmit power that limits its ASR performance. Further, we see that at high $P_{ave}$, e.g., $P_{ave}=10$ dBm, Alice's maneuverability has less impact on the ASR for $\lambda=0.5$. Also, the ASR performance for AN-based schemes is sensitive to power ratio $\lambda$, and higher power ratio $\lambda=0.9$ for lower $P_{ave}$ values, yields best ASR. This implies that when the total transmit power is low, the higher fraction should be dedicated to Alice's transmission, while for higher amounts of $P_{tot}$, equal power allocation ($\lambda=0.5$) is appealing, to achieve the best ASR for \textit{JTDORA}.
\vspace{-5mm}
\section{Conclusions}\label{ConCl} 
In this paper, we presented a secure two-phase transmission  protocol  via AN-assisted UAV communications. To combat eavesdropping, we formulated an ASR optimization problem in terms of path planning and resource allocation, and then proposed an iterative algorithm to solve it. The proposed computationally efficient algorithm provides suboptimal solution in terms of network transmit power, UAV's flying trajectory, and AN injection factor in order to maximize the ASR. We evaluated our proposed scheme via simulations in terms of ASR and flying trajectory, and demonstrated its effectiveness. 

 \bibliographystyle{IEEEtran}
\typeout{}\bibliography{myReferences}

\begin{IEEEbiography}[{\includegraphics[width=1in,height=1.25in,clip,keepaspectratio]{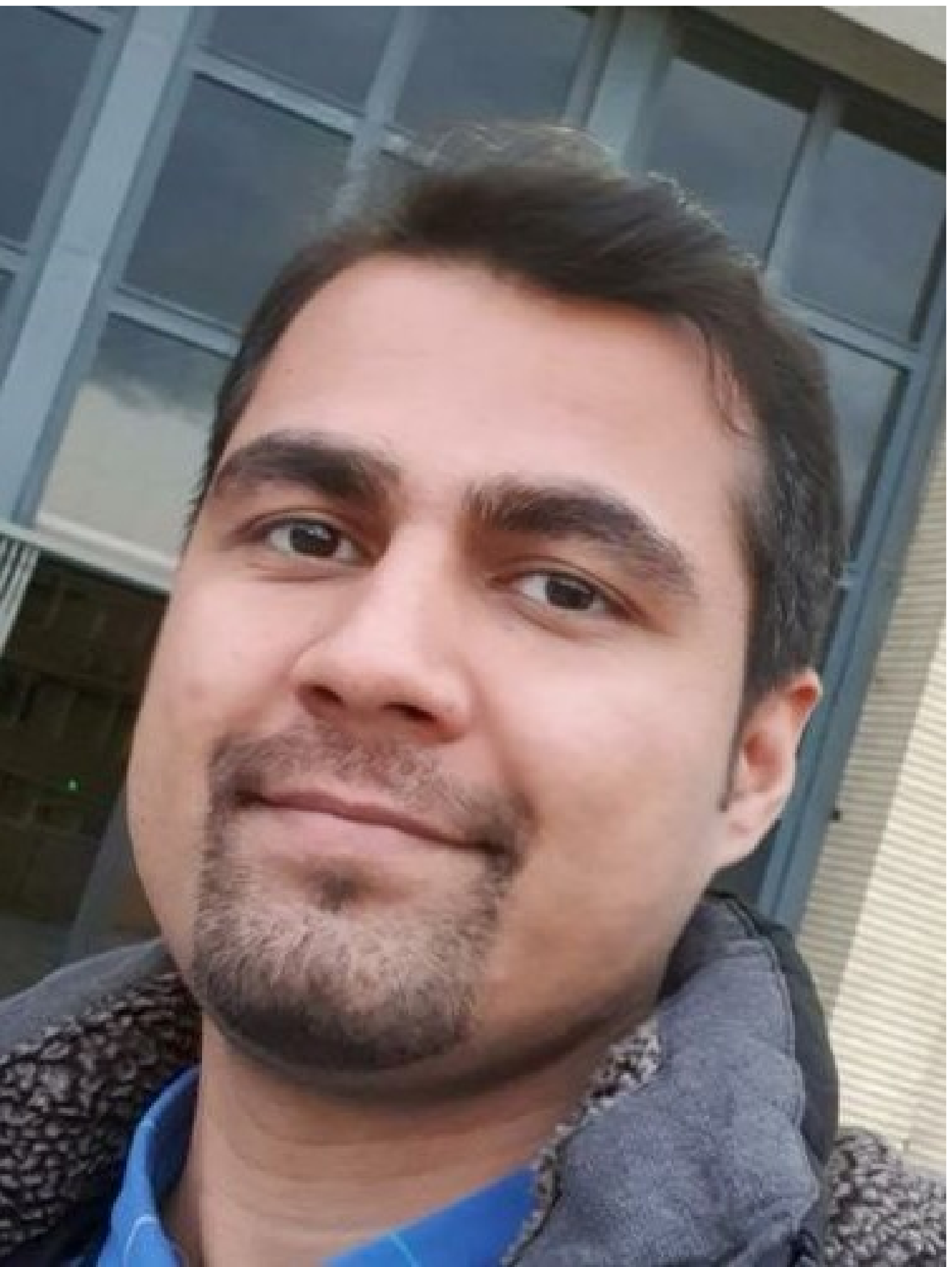}}]{Milad Tatar Mamaghani} was born in Tabriz, Iran, on May 12, 1994. He received the B.Sc. degrees in Electrical-Communications engineering (Major) and Control engineering (minor), from the Amirkabir University of Technology, Tehran, Iran. He is currently pursuing the Ph.D. degree with the Department of Electrical and Computer Systems Engineering, Monash University, Melbourne, Australia. Milad has served as a reviewer of many and various prestigious IEEE transactions and IET journals and conferences. His research interests mainly focus on B5G wireless communications and networking, Physical-layer security, UAV communications, and optimization.
\end{IEEEbiography}

\vskip 0pt plus -1fil

\begin{IEEEbiography}[{\includegraphics[width=1in,height=1.25in,clip,keepaspectratio]{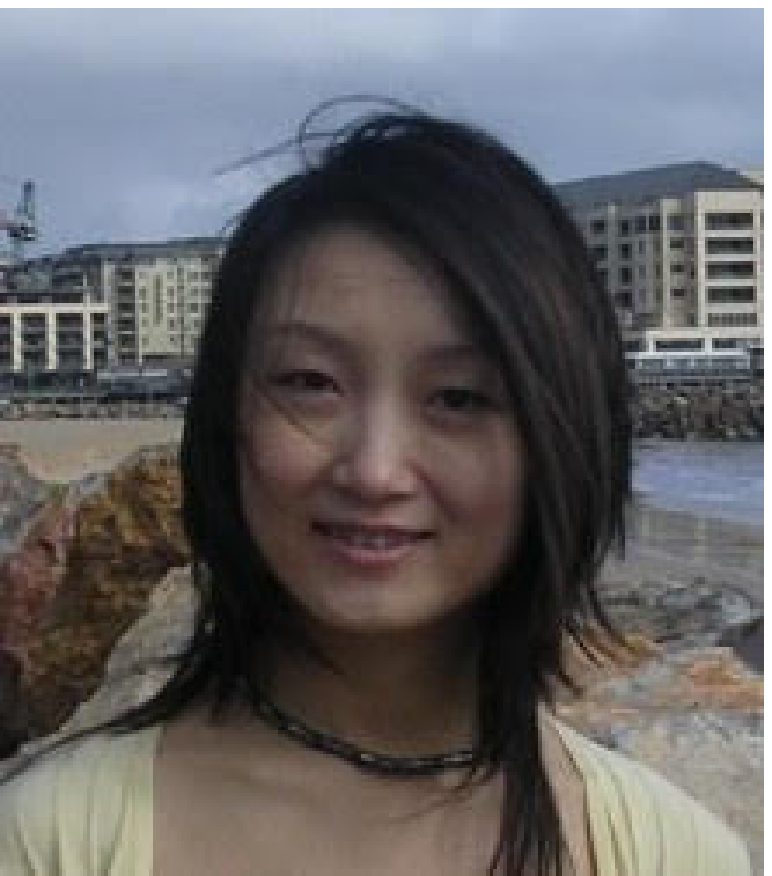}}]%
{Yi Hong}(S'00--M'05--SM'10)
is currently a Senior lecturer at the Department of Electrical and Computer Systems Eng.,
Monash University, Melbourne, Australia.
She obtained her Ph.D. degree in Electrical Engineering and Telecommunications 
from the University of New South Wales (UNSW), Sydney, and received   
the {\em NICTA-ACoRN Earlier Career Researcher Award} at the {\em Australian Communication
Theory Workshop}, Adelaide, Australia, 2007. She currently serves on the Australian Research Council College of Experts (2018-2020). Dr. Hong was an Associate Editor for {\em IEEE Wireless Communication Letters} and {\em Transactions on Emerging Telecommunications Technologies (ETT)}.
She was the General Co-Chair of {\em IEEE Information Theory Workshop} 2014, Hobart; the Technical Program Committee Chair of
{\em Australian Communications Theory Workshop} 2011, Melbourne; and the Publicity Chair
at the {\em IEEE Information Theory Workshop} 2009, Sicily. She was a Technical Program Committee member for
many IEEE leading conferences. Her research interests include
communication theory, coding and information theory with applications to telecommunication engineering.
\end{IEEEbiography}

\end{document}